\documentclass[10pt,journal]{IEEEtran}
\usepackage[utf8]{inputenc}
\usepackage[T1]{fontenc}
\usepackage{amsmath,amssymb,amsfonts,mathtools}
\usepackage{bm}
\usepackage{graphicx}
\usepackage{parskip}
\usepackage{tikz}
\usetikzlibrary{arrows.meta,positioning,calc}
\usepackage{booktabs}
\usepackage{siunitx}
\usepackage{orcidlink}
\usepackage{hyperref}
\usepackage{verbatim}
\usepackage[ruled,vlined]{algorithm2e}
\usepackage{xspace}
\usepackage{enumitem}
\usepackage{microtype}
\usepackage{balance}
\usepackage[backend=biber,style=ieee,maxbibnames=6]{biblatex}
\hypersetup{
  colorlinks=true,
  linkcolor=blue,
  citecolor=blue,
  urlcolor=blue,
  pdfauthor={Noureldin M. A. A. Mohamed},
  pdftitle={Adversarial Learning Game for Intrusion Detection in Quantum Key Distribution}
}
\newcommand{\FAR}{\mathrm{FAR}}
\newcommand{\AUC}{\(\mathrm{AUC}\)}

\title{Adversarial Learning Game for Intrusion Detection in Quantum Key Distribution}

\author{
    \IEEEauthorblockN{
        Noureldin Mohamed\IEEEauthorrefmark{1}\orcidlink{0009-0001-4150-8690} and 
        Saif Al-Kuwari\IEEEauthorrefmark{1}\orcidlink{0000-0002-4402-7710}
    }
    \vspace{0.2cm} \\ 
    \IEEEauthorblockA{
        \IEEEauthorrefmark{1}\textit{Qatar Center for Quantum Computing, College of Science and Engineering,} \\
        \textit{Hamad Bin Khalifa University, Doha, Qatar.} \\
        Corresponding author: nomo89098@hbku.edu.qa
    }
}
\begin{document}
\maketitle
\begin{abstract}
While Quantum Key Distribution (QKD) provides information-theoretic security, the transition from theory to physical hardware introduces side-channel vulnerabilities that traditional error metrics often fail to characterize. This paper presents a high-fidelity simulation framework for intrusion detection in decoy-state QKD, modeled as a minimax game between a learning-based defender and a physically constrained, adaptive adversary. The defender utilizes block-level telemetry (comprising decoy-state residuals, timing-histogram moments, and detector imbalances) to trigger alarms that gate key distillation . Unlike heuristic thresholds, our optimization objective is strictly operational: missed detections are penalized based on the resulting degradation of the finite-key secret fraction calculated via three-intensity decoy estimators and entropy-accumulation (EAT) penalties. 
The emulated adversary performs an automated search over time-shift, detector-blinding, photon number splitting (PNS), and Trojan-horse families, subject to hardware-limited feasibility bands. Concurrently, the defender co-trains one-class and temporal detectors (LSTM/TCN) using hard-negative mining to minimize the missed-attack rate at a calibrated false-alarm rate ($\text{FAR}$). Under adaptive attack scenarios, the system preserves $82\text{--}92\%$ of the honest finite-key rate while discarding only approximately $1.2\%$ of traffic, representing a net gain of $+20\text{--}35$ percentage points in usable secret bits over non-adversarial baselines. These results demonstrate that optimizing detection directly for secret-bit retention provides a robust, physically grounded layer of defense against adaptive side-channel strategies in practical QKD deployments.
\end{abstract}


\section{Introduction}
\label{sec:intro}

Quantum key distribution (QKD) promises information-theoretic security under explicit assumptions about device and channel behavior \cite{BB84,Scarani2009,Lo2014,GLKP2004}. In deployed systems, however, the transition from theoretical protocols to physical hardware introduces complex side-channel vulnerabilities. Eavesdroppers can exploit these device imperfections to compromise key material while leaving headline error metrics, such as the quantum bit error rate (QBER), largely undisturbed. Canonical examples of such physical layer attacks include:

\begin{itemize}
    \item \text{Time-shift attacks:} manipulating relative detection efficiencies by offsetting photon arrival times within the gating window \cite{Qi2007TS,Zhao2008TS}.
    \item \text{Detector blinding/control attacks:} driving single-photon avalanche diodes (SPADs) into the linear operating regime to deterministically control click statistics and suppress anomalous double-clicks \cite{Makarov2009SPD,Lydersen2010Blinding,Wiechers2011AfterGate}.
    \item \text{Photon-number splitting (PNS):} exploiting multi-photon signal components inherent in weakened coherent pulses under severe channel loss, allowing the adversary to extract information without triggering fundamental error alarms \cite{Brassard2000PNS,Hwang2003Decoy,Wang2005Decoy,Ma2005}.
    \item \text{Trojan-horse attacks (THA):} actively probing internal phase modulators and inferring basis settings via back-reflected light, bounded only by the sensitivity of internal optical monitors \cite{Jain2014THA,Jain2011Calib}.
\end{itemize}

While architectural countermeasures, such as measurement-device-independent QKD (MDI-QKD), effectively remove detector-side channels, they often incur higher experimental complexity, require stringent synchronization, and yield reduced key rates in specific distance regimes \cite{Lo2012MDI,Scarani2009,Lo2014}. Consequently, standard decoy-state BB84 deployments remain highly relevant but require robust, intelligent monitoring.

Crucially, physical side-channel attacks—such as time-shift exploits, detector blinding, and Trojan-horse probes—are deliberately engineered to evade standard error monitoring. Rather than causing pronounced spikes in the quantum bit error rate (QBER), they typically manifest as \emph{subtle, structured deviations} across multidimensional system telemetry \cite{Ma2005,Scarani2009}. These hidden signatures include decoy-state gain and error-gain residuals relative to Poisson-mixture predictions, timing-histogram skewness and kurtosis, per-detector click imbalances, and anomalous double-click rates. When left undetected, the resulting parameter drifts tighten single-photon bounds and enlarge finite-size penalties, ultimately eroding the \emph{finite-key} secret fraction that determines the operational throughput of the link \cite{Lo2014,EAT,Tomamichel2012FiniteKey}.

Despite the sophistication of these threats, current operational monitoring systems and the related intrusion-detection literature exhibit fundamental limitations. Traditional defense mechanisms typically deploy fixed, static thresholds in a handful of isolated statistics. However, these heuristic boundaries are rarely calibrated against a \emph{strategic} adversary. A rational attacker actively shapes their perturbation to remain strictly within the feasibility bands induced by natural hardware limits and decoy-state consistency, leaving highly exploitable blind spots in standard threshold-based defenses \cite{Qi2007TS,Lydersen2010Blinding,Jain2014THA,Fung2009Mismatch}.

More recent literature has explored machine learning and anomaly detection for QKD telemetry, yet these approaches largely fail to address the core operational requirements. First, generic anomaly detectors are predominantly trained on average-case channel noise or non-adversarial anomalies, ignoring the worst-case nature of cryptographic threats. Second, conventional ML models in this domain are tuned purely by statistical false-alarm considerations (e.g., standard cross-entropy loss) rather than optimizing the actual \emph{operational objective}: the retention of secret bits under rigorous finite-size effects. This fundamental mismatch leads to \emph{alarm fatigue}, where overly sensitive models discard excessive amounts of honest traffic and suppress usable throughput, or \emph{silent erosion} of extractable entropy when models fail to heavily penalize attacks that maximally degrade the secure key rate \cite{Scarani2009,Lo2014}. While recent advances in sequential detection and statistical risk control—such as conformal calibration—provide finite-sample error guarantees in classical machine learning \cite{Angelopoulos2023Conformal}, they have yet to be successfully integrated with rigorous decoy-state and entropy-accumulation (EAT) accounting to directly optimize secret-bit retention in quantum networks.

\subsection{Contributions}
In this work, we address these gaps through a simulation-driven adversarial learning framework. Our contributions can be summarized as follows:
\begin{itemize}
   \item \text{Operational, finite-key-aware objective:} we couple alarm triggers directly to secret-bit accounting. Missed detections incur a loss explicitly scaled to the degradation of the finite-key secret fraction ($r$) derived from three-intensity decoy bounds and EAT penalties.
    \item \text{Physically constrained attacker:} we instantiate emulated time-shift, blinding, PNS, and THA families with explicit feasibility projections (gate-offset bounds, safe-illumination envelopes, and decoy-consistency bands) grounded in realistic physical constraints.
    \item \text{Adversarial training with hard negatives:} we implement an alternating minimax training procedure that couples one-class and temporal detectors (LSTM/TCN) with an inner-loop adversary to systematically mine attack realizations that are physically viable yet hardest to detect.
    %\item \text{Robust simulation evidence:} across simulated \SI{50}{}\text{--}\SI{100}{\kilo\meter} fiber links and block sizes $N \in \{5{\times}10^4,\,2{\times}10^5,\,2{\times}10^6\}$, the detector sustains high discrimination power ($\text{AUC} \in [0.97, 0.997]$) and preserves $82\text{--}92\%$ of the honest finite-key rate under adaptive attacks.
\end{itemize}
Through comprehensive evaluations across simulated \SI{50}{}\text{--}\SI{100}{\kilo\meter} fiber links and block sizes $N \in \{5{\times}10^4,\,2{\times}10^5,\,2{\times}10^6\}$, we demonstrate the efficacy of these contributions. The proposed detector sustains high discrimination power ($\text{AUC} \in [0.97, 0.997]$) and preserves $82\text{--}92\%$ of the honest finite-key rate under adaptive attacks, validating that optimizing detection directly for secret-bit retention provides a robust defense for practical QKD deployments.

%\subsection{Organization}
The remainder of this paper is organized as follows. Section~\ref{sec:prelim} establishes the notation and reviews decoy-state fundamentals alongside the underlying game-theoretic concepts. Section~\ref{sec:threat} defines the threat model, detailing the adversary's capabilities and physical constraints. Section~\ref{sec:formulation} formalizes the minimax game and introduces the finite-key-aware operational loss function. Section~\ref{sec:algorithms} details the proposed intrusion detection framework, encompassing the hybrid defender architecture and the alternating minimax training procedure. Section~\ref{sec:sim} describes the high-fidelity simulation environment, telemetry synthesis, and evaluation metrics. Section~\ref{sec:results} presents a comprehensive analysis of the experimental results, including discrimination performance, operational key retention, and system overhead. Finally, Section~\ref{sec:conclusion} concludes the paper with a discussion on current limitations and avenues for future research.

\section{Preliminaries}
\label{sec:prelim}
This section establishes the notation used throughout this paper and reviews the fundamental decoy-state and finite-key ingredients utilized in our framework.  In the context of our high-fidelity simulation, we distinguish (i) \emph{gains} and \emph{error gains}, which are the central observables in decoy analysis \cite{Ma2005,Scarani2009,Lo2014}, from (ii) blockwise empirical QBERs used only as sanity checks.

\subsection{Notation}
Table~\ref{tab:notation} summarizes the key mathematical symbols and variables used throughout the decoy-state analysis and finite-key accounting in this framework.

\begin{table}[htbp]
\caption{Summary of Key Notation}
\label{tab:notation}
\centering
% Changed the second column to a 'p' column to allow text wrapping
\begin{tabular}{@{}l p{0.65\columnwidth}@{}}
\toprule
\textbf{Symbol} & \textbf{Description} \\ \midrule
$\mu \in \{\mu_s, \mu_w, \mu_v\}$ & Optical intensities (signal, weak decoy, vacuum) \\
$B \in \{Z, X\}$ & Measurement bases (key and test bases) \\
$N$ & Total number of emitted pulses per aggregation block \\
$Q(\mu,B)$ & Empirical gain for intensity $\mu$ and basis $B$ \\
$E(\mu,B)Q(\mu,B)$ & Empirical error gain for intensity $\mu$ and basis $B$ \\
$Y_n^{B}$, $e_n^{B}$ & $n$-photon yield and error rate in basis $B$ \\
$\underline{Y}_1^{Z}$, $\overline{e}_1^{X}$ & Bounded single-photon yield ($Z$) and phase error ($X$) \\
$N_{s}^{Z}$ & Number of sifted signal emissions in basis $Z$ \\
$s_{1}^{Z,\mathrm{L}}$ & Lower bound on single-photon signal counts in basis $Z$ \\
$e_{1}^{X,\mathrm{U}}$ & Upper bound on single-photon phase error \\
$\varepsilon_{\mathrm{decoy}}$ & Tail probability / confidence parameter for decoy bounds \\
$\varepsilon$ & Overall security parameter ($\varepsilon_{\mathrm{EC}} + \varepsilon_{\mathrm{PE}} + \varepsilon_{\mathrm{PA}} + \varepsilon_{\mathrm{EAT}}$) \\
$r$ & Per-emitted-pulse secret key fraction \\
$r_0, r(a)$ & Secret fraction under honest operation ($r_0$) and attack ($r(a)$) \\
$\lambda_{\mathrm{EC}}$ & Error-correction leakage (including verification costs) \\
$\Delta_{\mathrm{EAT}}(\varepsilon, N)$ & Finite-size penalty term from entropy accumulation \\
$x = \Phi(\cdot)$ & High-dimensional block-level telemetry feature vector \\
\bottomrule
\end{tabular}
\end{table}

\subsection{BB84 with Three-Intensity Decoys}
In our simulation, Alice selects an optical intensity \(\mu\in\{\mu_s,\mu_w,\mu_v\}\) (signal, weak decoy, vacuum) and basis \(B\in\{Z,X\}\), prepares a phase-randomized coherent state whose photon number
\begin{equation}
\Pr[N=n\mid \mu] \;=\; e^{-\mu}\,\frac{\mu^n}{n!},\qquad n\in\mathbb{N}_0,
\end{equation}
and sends it to Bob over a lossy/noisy channel. For each emitted pulse, Bob records a detection event and (when applicable) an error. Aggregating by \((\mu,B)\) over a block of \(N\) pulses yields the empirical gain and error gain
\begin{equation}
\begin{split}
Q(\mu,B) &= \frac{\text{\# detections at }(\mu,B)}{\text{\# emissions at }(\mu,B)}, \\
E(\mu,B)\,Q(\mu,B) &= \frac{\text{\# errors at }(\mu,B)}{\text{\# emissions at }(\mu,B)}.
\end{split}
\label{eq:prelim:gain-error-gain}
\end{equation}
For clarity, we occasionally report the conditional error fraction \(E(\mu,B)\in[0,1]\), noting that the error gain is the product \(E(\mu,B)Q(\mu,B)\).

\subsection{Yields, Error Rates, and Decoy-State Constraints}
Let \(Y_n^{B}\) be the \(n\)-photon \emph{yield} (detection probability) in basis \(B\), and \(e_n^{B}\) the corresponding error rate. Under phase randomization, the Poisson mixture gives the standard decoy-state relationships \cite{Ma2005,Scarani2009}:
\begin{equation}
\begin{split}
Q(\mu,B) &= \sum_{n=0}^{\infty} e^{-\mu}\frac{\mu^n}{n!}\, Y_n^{B}, \\
E(\mu,B)\,Q(\mu,B) &= \sum_{n=0}^{\infty} e^{-\mu}\frac{\mu^n}{n!}\, e_n^{B} Y_n^{B}.
\end{split}
\label{eq:prelim:decoy-mixture}
\end{equation}
With three intensities \(\{\mu_s,\mu_w,\mu_v\}\), linear constraints derived from \eqref{eq:prelim:decoy-mixture} (with finite-sample confidence intervals) produce bounds on the single-photon parameters that drive security:
a lower bound on the single-photon yield in the key basis \(Z\),
\(\underline{Y}_1^{Z}(\varepsilon_{\mathrm{decoy}})\),
and an upper bound on the single-photon phase error in the test basis \(X\),
\(\overline{e}_1^{X}(\varepsilon_{\mathrm{decoy}})\)
\cite{Ma2005,Scarani2009,Lo2014}. We will use the convenient counts form
\begin{equation}
s_{1}^{Z,\mathrm{L}} \;\triangleq\; N_{s}^{Z}\,\underline{Y}_{1}^{Z}(\varepsilon_{\mathrm{decoy}}),
\qquad
e_{1}^{X,\mathrm{U}} \;\triangleq\; \overline{e}_{1}^{X}(\varepsilon_{\mathrm{decoy}}),
\label{eq:prelim:single-photon-bounds}
\end{equation}
where \(N_{s}^{Z}\) is the number of \emph{signal} emissions in basis \(Z\) that survive sifting in the block, and \(\varepsilon_{\mathrm{decoy}}\) controls the tail probability of the decoy estimator (e.g., via multiplicative Chernoff or Clopper–Pearson intervals).

\subsection{Finite-Key Secret Fraction (EAT-Aware)}
Let \(\lambda_{\mathrm{EC}}\) denote the error-correction leakage, including verification costs. We employ entropy accumulation (EAT) to rigorously account for finite-size effects \cite{EAT,Tomamichel2012FiniteKey}. The per-emitted-pulse secret fraction is bounded by:
\begin{equation}
r \;\ge\; \frac{1}{N}\!\left(
s_{1}^{Z,\mathrm{L}}\big[1-h\!\left(e_{1}^{X,\mathrm{U}}\right)\big]
-\lambda_{\mathrm{EC}}
-\Delta_{\mathrm{EAT}}(\varepsilon,N)
\right),
\label{eq:prelim:finite-key}
\end{equation}
where \(h(\cdot)\) is the binary entropy function and \(\Delta_{\mathrm{EAT}}(\varepsilon,N)=\tilde{\mathcal{O}}\!\big(\sqrt{\log(1/\varepsilon)/N}\big)\) aggregates the finite-size security penalties. The overall security parameter is allocated as \(\varepsilon=\varepsilon_{\mathrm{EC}}+\varepsilon_{\mathrm{PE}}+\varepsilon_{\mathrm{PA}}+\varepsilon_{\mathrm{EAT}}\) with a fixed total budget (e.g., \(\varepsilon\le 10^{-10}\)) used consistently across all reports. We denote by \(r_0\) the value of \eqref{eq:prelim:finite-key} under honest operation (no attack) and by \(r(a)\) the value under an attack parameterization \(a\).

\subsection{Telemetry and Block-Level Feature Map}
Beyond the sufficient statistics for decoy estimation, we construct a low-dimensional feature vector \(x=\Phi(\cdot)\in\mathbb{R}^d\) per block intended for detection and attribution:
\begin{itemize}[leftmargin=*,nosep]
  \item \emph{Decoy residuals:} norms and signed components of the deviations between the observed \(\{Q(\mu,B),E(\mu,B)Q(\mu,B)\}\) and their best-fit Poisson-mixture reconstructions from \eqref{eq:prelim:decoy-mixture}.
  \item \emph{Timing structure:} histogram skewness/kurtosis of detection times, inter-gate asymmetry, or low-order moments of fitted Gaussian mixtures (sensitive to time-shift and after-gate phenomena) \cite{Qi2007TS,Wiechers2011AfterGate}.
  \item \emph{Detector signatures:} per-detector click imbalance and double-click rate (relevant to blinding/control attacks) \cite{Lydersen2010Blinding,Makarov2009SPD}.
  \item \emph{Environmental proxies:} if available, summaries of bias currents or temperature (to diagnose illumination envelopes and slow drifts).
\end{itemize}
These features are chosen to be physically interpretable, sensitive to known attack families, and statistically stable for moderate block sizes.

\subsection{Game-Theoretic Foundations}
\label{subsec:prelim:game_theory}

We briefly introduce the game-theoretic concepts that underpin our approach. In standard machine learning, a model is typically trained to minimize errors over a static, average-case dataset. However, in a cryptographic security context, the environment is actively hostile. We therefore model the interaction between the intrusion detection system and the eavesdropper as a two-player, strictly competitive game.

Let the defender, denoted as $\mathcal{D}$, be parameterized by a set of model weights and detection thresholds $\theta$. The adversary, $\mathcal{A}$, selects an attack configuration $a$ from a bounded, physically constrained feasible set $\mathbb{A}$. 

The interaction is governed by an operational loss function, $\ell(\theta; a)$, which quantifies the overall degradation of the quantum link. This loss captures both the reduction in the finite-key secret fraction caused by an active attack and the operational overhead incurred by false alarms on honest traffic. 

Because the attacker's goal (maximizing disruption and information leakage) is diametrically opposed to the defender's goal (maximizing secure key throughput), this dynamic represents a zero-sum arm race. The adversary systematically searches for the worst-case perturbation $a$ that inflicts maximum damage without triggering an alarm. Conversely, the defender updates their parameters $\theta$ to minimize this exact worst-case damage. 

Mathematically, this is expressed as a minimax optimization problem:
\begin{equation}
\min_{\theta} \max_{a \in \mathbb{A}} \ell(\theta; a).
\label{eq:prelim:minimax}
\end{equation}
By solving this minimax objective, the defender is forced to learn robust, physically grounded invariants of the quantum channel rather than memorizing specific attack patterns. This ensures that the system's security guarantees hold even when the adversary dynamically adapts their strategy to bypass the current detection rules.

\section{Threat Model \& Assumptions}
\label{sec:threat}

We consider a decoy-state BB84 protocol operated over a simulated lossy fiber channel \cite{Scarani2009,Lo2014,Ma2005}. The system is modeled as a repeated zero-sum game played at the level of aggregation blocks. An adaptive adversary ($\mathcal{A}$) manipulates the quantum channel and detector responses within a physically feasible set $\mathbb{A}$, while the defender ($\mathcal{D}$) observes block-level telemetry to raise alarms that gate key distillation.

\subsection{Adversary Capabilities and Constraints}
\label{subsec:attacker_constraints}
We instantiate four representative attack families grounded in experimental literature. Crucially, the adversary is not omnipotent; their actions are projected onto a feasible set $\mathbb{A} = \bigcup_f \mathbb{A}_f$ defined by hardware and monitoring limits.

%\subsubsection{Attacks}
\begin{itemize}
    \item \text{Time-shift ($\Delta t$):} the adversary induces detection efficiency mismatches by applying an intentional arrival offset $\Delta t$ relative to the detector gating window \cite{Qi2007TS,Zhao2008TS}.
    \item \text{Detector control / Blinding ($I(t)$):} The adversary injects tailored bright illumination $I(t)$ to drive SPADs into linear mode, thereby altering click statistics and suppressing double-click events \cite{Makarov2009SPD,Lydersen2010Blinding,Wiechers2011AfterGate}.
    \item \text{Photon-number splitting (PNS, $f_{\text{split}}$):} the adversary performs preferential blocking or splitting of multi-photon signal components ($n \ge 2$) under channel loss, constrained by the necessity to maintain decoy-state gain statistics \cite{Brassard2000PNS,Hwang2003Decoy,Wang2005Decoy,Ma2005}.
    \item \text{Trojan-horse (THA, $\rho, P_{\text{ret}}$):} the adversary actively probes the modulators, inferring basis settings from back-reflected light with correlation $\rho$ and return power $P_{\text{ret}}$, limited by the sensitivity of internal leakage monitors \cite{Jain2014THA,Jain2011Calib}.
\end{itemize}

%\subsubsection{Feasibility Projections}
To ensure the simulation remains realistic, every attack candidate $a$ generated by the inner optimizer is projected onto the feasible set $\mathbb{A}$ via the following constraints:
\begin{itemize}
    \item \text{Timing Bound:} $|\Delta t| \le \Delta t_{\max}$, determined by the physical gate width and electronic jitter \cite{Qi2007TS}.
    \item \text{Safe Illumination:} the blinding envelope must satisfy $\|I\|_{\infty} \le I_{\max}$ (vendor-safe limits) and remain within bounded deviations for bias-current proxies \cite{Lydersen2010Blinding}.
    \item \text{Decoy Consistency:} for all intensities and bases $(\mu, B)$, the attacked gains and error gains must lie within the finite-sample confidence bands of the Poisson-mixture model at confidence level $\varepsilon_{\text{decoy}}$ \cite{Ma2005,Scarani2009}.
    \item \text{Leakage Monitor:} THA return power is strictly bounded by $P_{\text{ret}} \le P_{\max}$ and correlation $\rho \le \rho_{\max}$, simulating the thresholds of passive hardware monitors \cite{Jain2014THA}.
\end{itemize}

\subsection{Information Structure and Play}
We assume block-synchronous play. In each block, $\mathcal{A}$ selects a parameterization $a \in \mathbb{A}$ (potentially randomized). The defender computes a score $s_{\theta}(x)$ based on the resulting telemetry and emits an alarm $\pi_{\theta}(x) \in \{0,1\}$. Alarms gate the sifting and distillation processes for that specific block. 
Neither party observes the other's internal randomness within a block. However, across blocks, both agents may adapt their strategies based on the history of alarms (a repeated game with observed outcomes).

\subsection{Trust, Randomness, and Synchronization}
Our simulation assumes authenticated classical communications and private, trusted randomness for Alice and Bob's basis and intensity choices. We assume calibrated clocks and gating operations within vendor specifications \cite{Jain2014THA}. 
We do \emph{not} assume device independence; detector-side vulnerabilities are explicitly treated as part of the attack surface $\mathbb{A}$. Note that while MDI-QKD \cite{Lo2012MDI} can remove detector side channels, this work focuses on securing the widely deployed prepare-and-measure decoy-state architecture.

\section{Game-Theoretic Formulation \& Operational Loss}
\label{sec:formulation}

We formalize intrusion detection in decoy-state QKD as a two-player zero-sum game played within a high-fidelity simulation environment. The interaction is modeled as a repeated game in which the defender ($\mathcal{D}$) attempts to distinguish honest channel noise from strategic perturbations introduced by a physically constrained adversary ($\mathcal{A}$).

%\subsection{Pipeline Overview}
\label{subsec:pipeline}
The end-to-end interaction flow operates as follows for each aggregation block:
\begin{enumerate}
    \item \text{Attack Selection:} the adversary selects a perturbation parameter $a$ from the feasible set $\mathbb{A}$ (defined in Sec.~\ref{sec:threat}).
    \item \text{Emulated Telemetry:} the simulation generates block-level telemetry vector $x(a)$, including decoy residuals, timing moments, and detector statistics, based on the channel model and attack parameters.
    \item \text{Defender Scoring:} the defender computes an anomaly score $s_{\theta}(x(a))$ using a parameterized model (e.g., LSTM/TCN) and applies a calibrated threshold $\tau$ to make a binary decision $\pi_{\theta}(x) \in \{0,1\}$ (Pass/Alarm).
    \item \text{Operational Impact:} if an alarm is raised ($\pi=1$), the block is discarded. If the block passes ($\pi=0$), it contributes to the secret key. Crucially, the quality of this contribution is priced by the \emph{finite-key secret fraction} $r(a)$ computed via the decoy-state and EAT bounds described in Sec.~\ref{sec:prelim}.
\end{enumerate}

\begin{figure*}[t]
  \centering
  
  \includegraphics[width=0.95\textwidth]{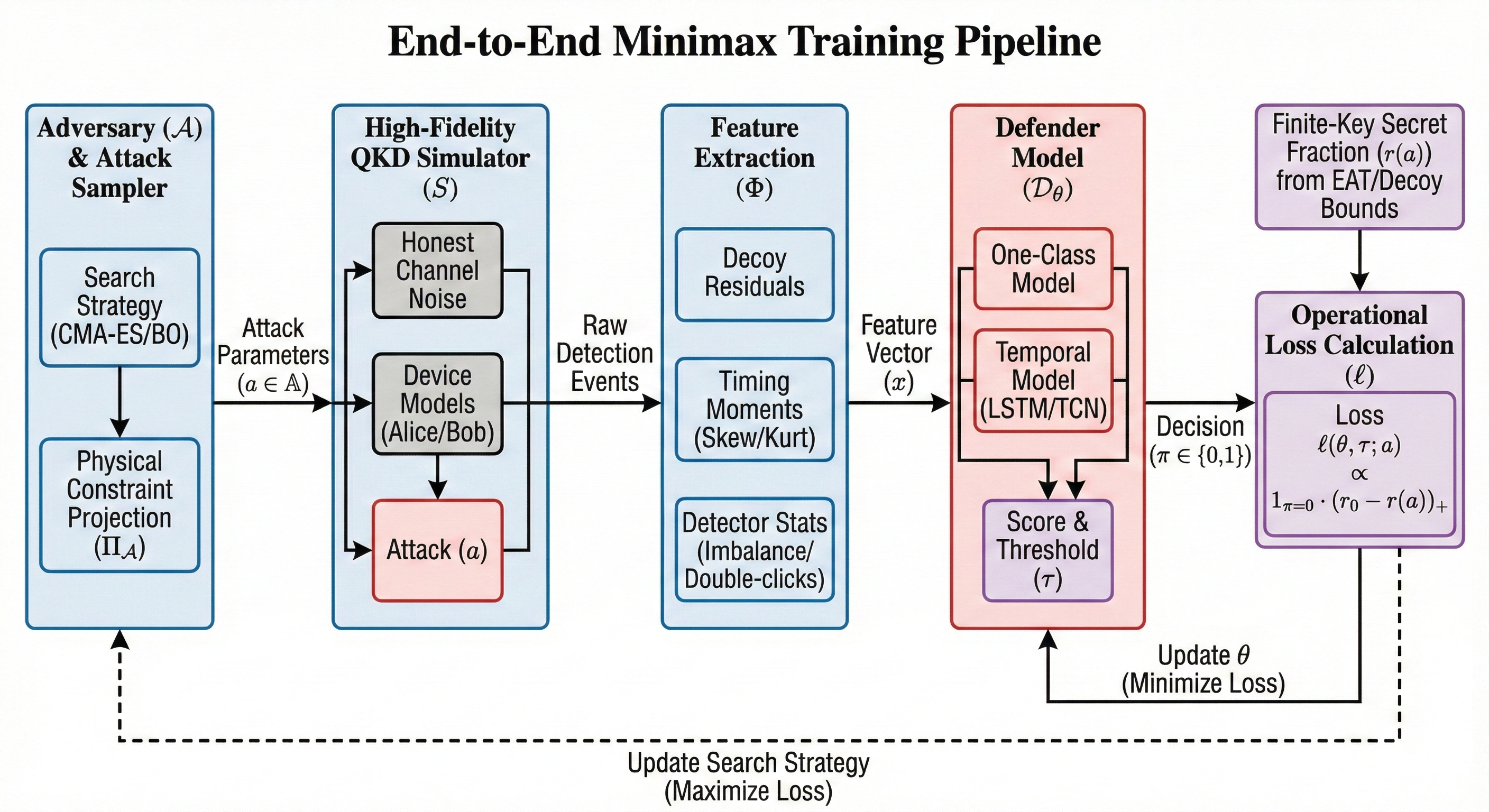}
  \caption{End-to-End Minimax Training Pipeline. The diagram captures the adversarial feedback loop: the adversary (blue) samples physically feasible attack parameters to maximize the operational loss, while the defender (red) updates its neural weights to minimize the degradation of the finite-key secret fraction ($r$). The simulator serves as the non-differentiable environment bridging the two agents.}
  \label{fig:minimax}
\end{figure*}
Figure~\ref{fig:minimax} details the computational flow of the training procedure. The cycle initiates with the \text{Adversary (\(\mathcal{A}\))}, which utilizes a derivative-free optimizer (CMA-ES or Bayesian Optimization) to propose attack parameters \(a\). These candidates are strictly projected onto the feasible set \(\mathbb{A}\) before being fed into the \text{High-Fidelity QKD Simulator}. The simulator generates raw detection events reflecting both the honest channel noise and the specific attack physics. This raw data is transformed into the feature space \(\Phi\) and processed by the \text{Defender (\(\mathcal{D}_\theta\))}, which outputs a binary decision \(\pi \in \{0,1\}\). 

The critical feedback mechanism is the \text{Operational Loss Calculation}. Unlike standard cross-entropy losses, our objective function weights missed detections by the \emph{realized security impact}: the difference between the honest secret fraction \(r_0\) and the attacked secret fraction \(r(a)\), calculated via the finite-key EAT bounds. This scalar loss signal \(\ell(\theta, \tau; a)\) is used to update the defender's parameters \(\theta\) via gradient descent (minimization) and to guide the adversary's search distribution toward more damaging regions of the attack space (maximization).
This pipeline explicitly couples the intrusion detection logic to the system's operational objective: maximizing the secure key rate. The training process alternates between finding the "hardest" physically feasible attacks (inner maximization) and updating the defender to mitigate secret-bit loss (outer minimization).

\subsection{Players, Strategies, and Information}
%\subsubsection{Attacker ($\mathcal{A}$)}
In each block, the adversary chooses $a \in \mathbb{A}$, where $\mathbb{A}$ is the union of feasible parameterizations for time-shift, blinding, PNS, and THA families subject to the hardware constraints detailed in Sec.~\ref{subsec:attacker_constraints}. Across blocks, $\mathcal{A}$ may randomize strategy and adapt to observed alarm rates to evade detection while maximizing information leakage.

%\subsubsection{Defender ($\mathcal{D}$)}
The defender commits to a parametric scoring rule $s_{\theta}: \mathbb{R}^{d} \to \mathbb{R}$ acting on the feature vector $x = \Phi(\cdot)$. Given a threshold $\tau$, the alarm policy is defined as:
\begin{equation}
\pi_{\theta}(x) = \mathbb{1}\{s_{\theta}(x) \ge \tau\}.
\end{equation}
We denote by $x(a)$ the random feature vector generated when the block is subjected to attack $a$, and by $x(\emptyset)$ the vector under honest operation.

\subsection{Finite-Key-Aware Defender Loss}
To ensure the defender optimizes for operational throughput rather than abstract classification metrics, we construct a loss function directly coupled to the finite-key secret fraction derived in Eq.~\eqref{eq:prelim:finite-key}.
Let $r_0$ denote the finite-key secret fraction under honest operation, and let $r(a)$ be the value under attack $a$, both computed using the same $\varepsilon$-security budget and three-intensity decoy bounds \cite{Ma2005,Lo2014,EAT}. The per-block loss function is defined as:
\begin{equation}
\begin{split}
l(\theta,\tau;a) \;=\; 
&\underbrace{\beta \cdot \mathbb{1}\{\pi_{\theta}(x(\emptyset))=1\}}_{\text{False Alarm Penalty}} \\
&+\; \underbrace{\mathbb{1}\{\pi_{\theta}(x(a))=0\}}_{\text{Missed Detection}} \cdot \left( \alpha + \gamma (r_0 - r(a))_+ \right).
\end{split}
\label{eq:formulation:loss}
\end{equation}
%\begin{itemize}
    %\item $\beta$ penalizes false alarms (discarding honest keys).
    %\item $\alpha$ provides a base penalty for any missed attack.
    %\item $\gamma$ scales the penalty proportional to the \emph{operational erosion} of the secret key, $(r_0 - r(a))_+$.
%\end{itemize}
In this formulation, $\beta$ penalizes false alarms (discarding honest keys), $\alpha$ provides a base penalty for any missed attack, and $\gamma$ scales the penalty proportional to the \emph{operational erosion} of the secret key, $(r_0 - r(a))_+$. Consequently, attacks that significantly degrade the secure key rate $r(a)$ incur a proportionally higher loss, forcing the defender to prioritize dangerous, high-impact exploits over negligible perturbations.

\subsection{Minimax and Stackelberg Objectives}
The training objective is to minimize the expected loss under the worst-case attack strategy. Formally, we seek the equilibrium:
\begin{equation}
\min_{\theta, \tau} \max_{a \in \mathbb{A}} \mathbb{E} \left[ l(\theta, \tau; a) \right],
\label{eq:formulation:minimax}
\end{equation}
where the expectation is taken over channel noise and sampling variability. 
In practice, we adopt a Stackelberg perspective: for fixed defender parameters $(\theta, \tau)$, we approximate the inner maximization $a^*(\theta, \tau) \in \arg\max_{a \in \mathbb{A}} \mathbb{E}[l]$ using a derivative-free optimizer (CMA-ES/Bayesian Optimization), and subsequently update $\theta$ to suppress the loss induced by $a^*$.

\subsection{Operating Point: FAR-Calibrated Thresholds}
To align with service-level agreements in deployed systems, we do not optimize $\tau$ freely. Instead, we select $\tau = \tau(\text{FAR})$ such that the false-alarm probability under honest operation meets a prescribed target (e.g., $\text{FAR}=1\%$):
\begin{equation}
\Pr[s_{\theta}(x(\emptyset)) \ge \tau(\text{FAR})] \approx \text{FAR}.
\label{eq:formulation:far}
\end{equation}
Calibration is performed on a held-out set of honest simulation blocks. Consequently, the optimization problem reduces to minimizing the \emph{missed-attack impact} (weighted by secret-bit loss) at a fixed operational FAR.

\section{Proposed Framework}
\label{sec:algorithms}

In this section, we detail the computational framework governing the defender's detection logic and the adversary's optimization loop. We describe the hybrid neural architecture, the projection-based attack search, and the alternating minimax training protocol. Furthermore, we outline advanced modules for sequential detection, distributional robustness, and bi-level decoy co-design. All finite-key security quantities follow the definitions in Sections~\ref{sec:prelim}--\ref{sec:formulation}.

\begin{figure*}[t]
  \centering
  \includegraphics[width=0.95\textwidth]{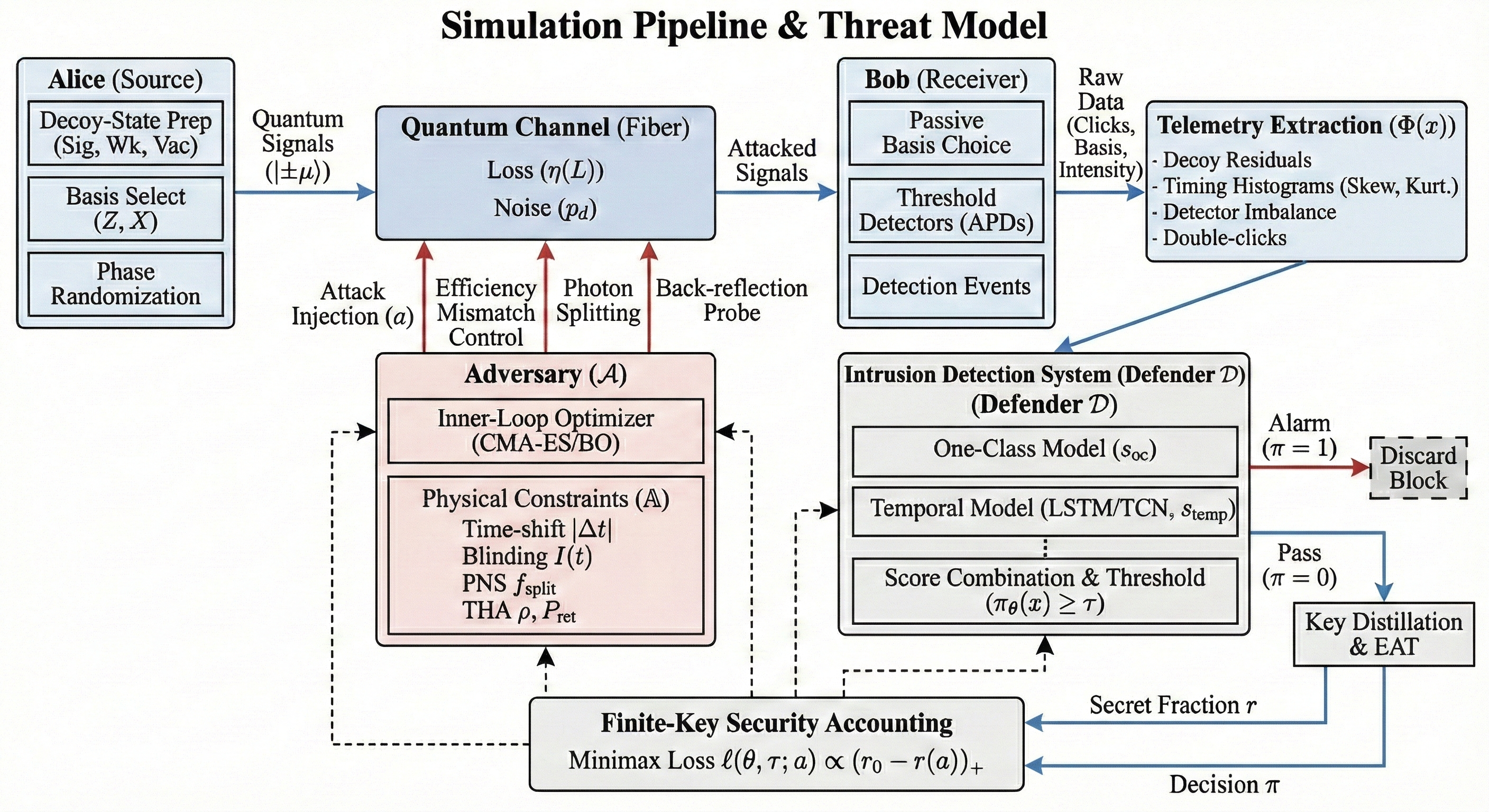}
  \caption{Schematic of the Simulation Pipeline \& Threat Model. The framework models the physical layer (Alice to Bob), the adversarial intervention (Red), and the defensive telemetry processing (Blue). Dashed lines indicate the gradient-free feedback loop used to train the adversary and the finite-key-aware loss used to update the defender.}
  \label{fig:pipeline}
\end{figure*}

Figure~\ref{fig:pipeline} illustrates the end-to-end architecture of our high-fidelity simulation framework. The pipeline begins with \text{Alice}, who generates phase-randomized decoy states according to the protocol specifications. These signals traverse a lossy quantum channel where the \text{Adversary (\(\mathcal{A}\))} injects perturbations---such as time-shifts (\(\Delta t\)), blinding envelopes (\(I(t)\)), or Trojan-horse probes---subject to the strict physical constraints defined in the feasible set \(\mathbb{A}\) (e.g., maximum jitter \(\Delta t_{\max}\) or safe illumination limits). At the receiver \text{Bob}, raw detection events are aggregated to compute the standard decoy statistics alongside a high-dimensional \text{Telemetry Vector (\(\Phi(x)\))}, comprising residuals, timing moments, and detector imbalances. The \text{Intrusion Detection System (Defender \(\mathcal{D}\))} processes this telemetry via parallel One-Class and Temporal (LSTM/TCN) models to generate an anomaly score. Crucially, the system closes the learning loop via the \text{Finite-Key Security Accounting} module: the defender's decision gates the key distillation process, and the resulting secret-key fraction \(r\) serves as the payoff function. This couples the updates of both the neural network weights and the adversary's search strategy directly to the operational throughput of the link.

\subsection{Defender Architecture}
\label{subsec:alg:defender}
To ensure sensitivity to both instantaneous violations and slow-moving trends, we employ a hybrid scoring model that combines unsupervised anomaly detection with a supervised temporal network.

%\subsubsection{Hybrid Scoring Mechanism}
The defender computes a score $s_\theta(x_t, \bm{x}_{t-w+1:t})$ for the current block $t$ using a convex combination of two components:
\begin{enumerate}
    \item \text{One-Class Anomaly Score ($s_{\mathrm{oc}}$):} An unsupervised model (e.g., Isolation Forest or One-Class SVM) trained exclusively on honest data. It maps the instantaneous feature vector $x_t$ to a scalar score, flagging point anomalies that deviate from the nominal manifold.
    \item \text{Temporal Score ($s_{\mathrm{temp}}$):} A sequence model (e.g., LSTM or TCN) that processes a sliding window of recent feature vectors $\bm{x}_{t-w+1:t} = (x_{t-w+1}, \dots, x_t)$. This component is trained to detect structured drifts and temporal correlations characteristic of adaptive attacks (e.g., blinding pulse sequences or gradual time-shifts).
\end{enumerate}

The final detection score is given by:
\begin{equation}
s_\theta(x_t,\bm{x}_{t-w+1:t}) = \lambda s_{\mathrm{oc}}(x_t) + (1-\lambda) s_{\mathrm{temp}}(\bm{x}_{t-w+1:t})
\label{eq:alg:mix}
\end{equation}
where $\lambda \in [0,1]$ is a hyperparameter tuned on a validation set to maximize AUC at the target operating point.

% [Placeholder: Figure 3 - Neural Network Architecture Diagram]
% Showing the parallel processing of One-Class and Temporal branches merging into the final score.

\subsection{Attacker Optimizer with Feasibility Projections}
\label{subsec:alg:attacker}
The inner loop of our training process solves for the "hardest" physically feasible attack $a^\star$. This is formulated as a constrained maximization of the defender's loss $\ell(\theta, \tau; a)$.

Since the gradient $\nabla_a \ell$ is generally intractable due to the complex channel simulation, we employ a derivative-free optimization strategy (e.g., CMA-ES or Bayesian Optimization). Crucially, every candidate attack $\tilde{a}$ proposed by the search is mapped to the feasible set $\mathbb{A}$ via a projection operator $\Pi_{\mathcal{A}}(\cdot)$. This operator strictly enforces:
\begin{itemize}
    \item \text{Timing Constraints:} $|\Delta t| \le \Delta t_{\max}$.
    \item \text{Safe Illumination:} $\|I\|_\infty \le I_{\max}$ and consistency with bias-current proxies.
    \item \text{Decoy Consistency:} Ensuring observed gains/error gains remain within the $\varepsilon_{\mathrm{decoy}}$ confidence intervals of the Poisson model.
    \item \text{Leakage Bounds:} $P_{\mathrm{ret}} \le P_{\max}$ and $\rho \le \rho_{\max}$.
\end{itemize}
For budgeted scenarios, we modify the objective to $\max_{a \in \mathbb{A}} (\ell(\theta, \tau; a) - \lambda_{\text{cost}} c(a))$, where $c(a)$ penalizes high-magnitude perturbations.
To operationalize this inner loop, we train the overall system using an alternating procedure that mimics an arms race, as detailed in Algorithm~\ref{alg:minimax}. The process iterates between updating the defender to minimize loss (weighted by secret-bit retention) and employing the aforementioned projection-based attacker to maximize the loss via hard-negative mining.
%\subsection{Alternating Minimax Training}
%\label{subsec:alg:minimax}
%We train the system using an alternating procedure that mimics an arms race. The process iterates between updating the defender to minimize loss (weighted by secret-bit retention) and updating the attacker to maximize it.

\begin{algorithm}[t]
\DontPrintSemicolon
\caption{Alternating Minimax Training with Hard-Negative Mining}
\label{alg:minimax}
\KwIn{Simulator $\mathcal{S}$, feasible set $\mathcal{A}$, initial defender $\theta_0$, target $\text{FAR}$}
\KwOut{Trained model parameters $(\theta, \tau)$}
$k \leftarrow 0$\;
\While{not converged}{
  \tcp{Step 1: Data Generation}
  Draw honest blocks $\mathcal{D}^{\mathrm{hon}} \sim \mathcal{S}(\emptyset)$\;
  Draw seeded/hard-negative blocks $\mathcal{D}^{\mathrm{atk}}$ from previous rounds\;
  
  \tcp{Step 2: Defender Update}
  Train $\theta_k$ to minimize detection error on $\mathcal{D}^{\mathrm{hon}} \cup \mathcal{D}^{\mathrm{atk}}$\;
  Calibrate $\tau_k$ such that $\Pr(s_{\theta_k}(x) \ge \tau_k \mid \text{honest}) \approx \text{FAR}$\;
  
  \tcp{Step 3: Inner Maximization (Hard Negative Mining)}
  Initialize attack population $\mathcal{P}$\;
  \For{inner\_step $= 1$ \KwTo $T$}{
    Suggest candidates $\tilde{a} \sim \mathcal{P}$ and project $a \leftarrow \Pi_{\mathcal{A}}(\tilde{a})$\;
    Evaluate loss $\ell(\theta_k, \tau_k; a)$ via simulation\;
    Update $\mathcal{P}$ to maximize loss (e.g., CMA-ES update)\;
  }
  $a^\star_k \leftarrow$ best found attack; Generate new hard negatives $\mathcal{D}^{\mathrm{new}} \sim \mathcal{S}(a^\star_k)$\;
  
  \tcp{Step 4: Augmentation}
  $\mathcal{D}^{\mathrm{atk}} \leftarrow \mathcal{D}^{\mathrm{atk}} \cup \mathcal{D}^{\mathrm{new}}$\;
  $k \leftarrow k+1$\;
}
\Return $(\theta_k, \tau_k)$\;
\end{algorithm}

\subsection{Advanced Detection and Robustness Modules}
\label{subsec:alg:advanced}
To further enhance the operational resilience and responsiveness of the intrusion detection framework, we integrate four advanced auxiliary modules. These extensions are designed to minimize detection latency across continuous block streams, prevent the attacker from overfitting to a single worst-case strategy, provide certified mathematical guarantees against feature noise, and optimize the underlying quantum protocol parameters in tandem with the neural defender. The specific components are detailed as follows:
\paragraph{Sequential Detection (CUSUM)}
To reduce detection latency, we augment the blockwise decision with a Cumulative Sum (CUSUM) control chart. We fit likelihood ratios $L_t = \log(p_1(s_t)/p_0(s_t))$ based on the score distributions of honest ($p_0$) and attacked ($p_1$) blocks. An alarm is triggered at time $T = \min \{t : S_t \ge h\}$, where $S_t = \max(0, S_{t-1} + L_t)$ accumulates evidence of a persistent attack.

\paragraph{Distributional Robustness (DRO)}
To prevent the adversary from overfitting to a single "worst-case" parameter, we employ a Distributionally Robust Optimization (DRO) approach. Instead of finding a point maximizer $a^\star$, the inner loop learns a worst-case distribution $P$ over $\mathbb{A}$ by sampling mixtures of attack families and reweighting them via exponential utility, subject to a Wasserstein transport budget.

\paragraph{Randomized Smoothing}
We provide certified robustness guarantees against feature noise by smoothing the decision boundary. The smoothed score is estimated via Monte Carlo sampling: $\bar{s}_\theta(x) = \mathbb{E}_{\xi \sim \mathcal{N}(0, \sigma^2 I)} [s_\theta(x+\xi)]$. If the margin $|\bar{s}_\theta(x) - \tau|$ exceeds a concentration bound, the decision is certified invariant within a radius proportional to $\sigma$.

\paragraph{Bi-Level Decoy Co-Design}
Finally, we extend the optimization to the protocol level by co-designing the decoy schedule $\bm{p}_\mu$ with the detector. We differentiate through unrolled inner optimization steps to compute gradients $\nabla_{\bm{p}_\mu} \mathbb{E}[\ell]$, allowing the system to automatically discover decoy intensity distributions that maximize the distinguishability of attacks under an average power constraint.

%\subsection{Complexity and Runtime Analysis}
%\label{subsec:alg:complexity}
%Let $d$ be the feature dimension and $w$ the temporal window. The inference cost per block is dominated by the temporal model, scaling as $O(wd^2)$. The training overhead is driven by the inner maximization loop; searching a population $p$ over $T$ iterations incurs a cost of $O(pT \cdot C_{\mathcal{S}})$, where $C_{\mathcal{S}}$ is the cost of a block simulation. Empirically, we find that the alternating minimization stabilizes within $5$--$7$ rounds (see Sec.~\ref{sec:results}), making the training computationally tractable for emulated high-rate QKD systems.

\section{Evaluation}
\label{sec:sim}

We evaluate the proposed intrusion detection framework using a high-fidelity emulation of the BB84 protocol with three-intensity decoy states over a lossy, noisy fiber channel. This environment generates block-level observables and telemetry consistent with the physical behavior of practical devices. Unless stated otherwise, results are reported for fiber distances \(L\in\{\SI{50}{\kilo\meter},\,\SI{100}{\kilo\meter}\}\) and block sizes \(N\in\{5{\times}10^4,\,2{\times}10^5,\,2{\times}10^6\}\).

\subsection{Channel and Device Models}
\label{subsec:sim:channel}

%\paragraph{Loss and dark counts.}
The channel transmittance is modeled as \(\eta(L)=10^{-\alpha L/10}\) with a standard attenuation coefficient \(\alpha=\SI{0.2}{\decibel\per\kilo\meter}\). The dark-count probability per gate is modeled log-uniformly in the range \(p_d\in[10^{-7},\,10^{-6}]\), reflecting typical commercial InGaAs APD performance.
%
%\paragraph{Misalignment and visibility.}
Optical misalignment is modeled as a basis-symmetric error probability \(e_d\in[1.0\%,\,1.8\%]\), which induces a system visibility \(V=1-2e_d\). This contributes an intrinsic error floor of \(e_d/2\) per detected event.
%
%\paragraph{Photon-number yields and error rates.}
For a coherent state with mean photon number \(\mu\), the probability of emitting \(n\) photons is Poissonian. The \(n\)-photon yield \(Y_n^{B}\) and error rate \(e_n^{B}\) in basis \(B\in\{Z,X\}\) are given by \cite{Ma2005,Scarani2009}:
\begin{equation}
Y_n^{B} \;=\; 1-(1-\eta)^n + p_d, \qquad
e_n^{B} \;=\; \frac{1-V}{2} + \frac{p_d/2}{Y_n^B}.
\end{equation}
The error rate equation includes a background contribution that dominates when signal counts are low. The observable gain \(Q(\mu,B)\) and error gain \(E(\mu,B)Q(\mu,B)\) follow the standard Poisson mixing:
\begin{align}
Q(\mu,B) &= \sum_{n\ge0} e^{-\mu}\frac{\mu^n}{n!}\,Y_n^{B}, \label{eq:sim:gain}\\
E(\mu,B)\,Q(\mu,B) &= \sum_{n\ge0} e^{-\mu}\frac{\mu^n}{n!}\,e_n^{B}Y_n^{B}. \label{eq:sim:errgain}
\end{align}

\subsection{Decoy Settings and Basis Choice}
The protocol employs three intensities \(\mu\in\{\mu_s,\mu_w,\mu_v\}\) configured as follows: \(\mu_s\approx0.5\text{--}0.6\) (signal), \(\mu_w\approx0.1\) (decoy), and \(\mu_v\approx0\) (vacuum).
Selection probabilities are set to \(p(\mu)=(p_s,p_w,p_v)\) with \(p_s\in[0.6,0.8]\) and \(p_w\in[0.15,0.35]\). To maximize the generation rate, the key basis is favored with \(p(Z)\in[0.8,0.95]\), while \(p(X)=1-p(Z)\) is used strictly for parameter estimation.

\subsection{Telemetry and Feature Synthesis}
\label{subsec:sim:telemetry}
For each block, the simulation extracts a high-dimensional feature vector \(x=\Phi(\cdot)\in\mathbb{R}^d\) designed to expose side-channel signatures:
\begin{itemize}[leftmargin=*,nosep]
  \item \text{Decoy residuals:} The Euclidean norms and signed deviations between the observed statistics \(\{Q(\mu,B),E(\mu,B)Q(\mu,B)\}\) and the theoretical expectations derived from the best-fit Poisson mixture (Eqs.~\eqref{eq:sim:gain}--\eqref{eq:sim:errgain}).
  \item \text{Timing structure:} Statistical moments (skewness, kurtosis) and inter-gate asymmetry computed from detection-time histograms. Photon arrival times are sampled from a mixture-of-Gaussians model with a gate jitter \(\sigma_t\in[\SI{50}{\pico\second},\,\SI{120}{\pico\second}]\), sensitive to time-shift attacks \cite{Qi2007TS}.
  \item \text{Detector signatures:} The per-detector click imbalance and double-click rate, calculated from two simulated APD channels sharing \(\eta\) and \(p_d\). These are primary indicators for blinding attempts \cite{Lydersen2010Blinding}.
  \item \text{Environmental proxies:} (Optional) Simulated readouts of bias currents or temperature to account for slow environmental drifts.
\end{itemize}

\subsection{Attack Families and Feasibility Constraints}
\label{subsec:sim:attacks}
The adversary \(\mathcal{A}\) optimizes over four parameterized attack families. To ensure physical realism, all candidate parameters \(a\) are projected onto the feasible set \(\mathcal{A} = \bigcup_f \mathcal{A}_f\). The complete set of boundary conditions and underlying simulation parameters are summarized in Table~\ref{tab:simparams}:
\begin{itemize}[leftmargin=*,nosep]
  \item \text{Time-shift:} An arrival time offset \(\Delta t\in[-\Delta t_{\max},\,\Delta t_{\max}]\) is applied, where \(\Delta t_{\max}\in[\SI{80}{\pico\second},\,\SI{150}{\pico\second}]\). This induces an efficiency mismatch between detectors due to the shifted gating window \cite{Qi2007TS}.
  \item \text{Detector blinding:} A continuous illumination envelope \(I(t)=I_0\,\mathbf{1}_{[t_1,t_2]}\) is injected. The power is bounded by \(I_0\le I_{\max}\) (vendor-safe limit), and the attack is constrained by deviations in the double-click rate and bias-current proxies \cite{Lydersen2010Blinding}.
  \item \text{PNS family:} Multi-photon yields (\(n\ge2\)) are manipulated via a split fraction \(f_{\mathrm{split}}\in[0,1]\). The attack is constrained to maintain decoy-state gains and error gains within the statistical confidence intervals of the honest model \cite{Brassard2000PNS}.
  \item \text{Trojan-horse (THA):} The adversary probes with correlation \(\rho\in[0,\rho_{\max}]\) and return power \(P_{\mathrm{ret}}\le P_{\max}\), limited by the sensitivity of internal leakage monitors \cite{Jain2014THA}.
\end{itemize}

\subsection{Domain Randomization and Robustness}
To prevent the defender from overfitting to specific channel conditions, we apply domain randomization to every training block. We sample \(\eta\) within a \(\pm\SI{1}{\decibel}\) window, \(e_d\) within \(\pm0.4\) percentage points, and \(\sigma_t\) within \(\pm\SI{20}{\pico\second}\). Decoy probabilities are perturbed by \(\pm0.05\), and dark counts are sampled log-uniformly. This ensures the learned decision boundary is robust to natural device variations.

\subsection{Datasets, Splits, and Metrics}
We generate a multi-condition corpus split strictly by simulation seed and drift conditions to prevent data leakage. The rich telemetry extraction ensures strong distinguishability between nominal and malicious operations; for example, Figure~\ref{fig:channelmodels} visualizes the separability of the resulting feature distributions under honest and attacked conditions. :
\begin{itemize}[leftmargin=*,nosep]
  \item \text{Data Split:} \(60/20/20\) for Train/Validation/Test, stratified across \(\{L,N\}\) and drift settings.
  \item \text{Primary Metrics:} Area Under the ROC Curve (\AUC), Missed-Attack Probability (Miss@\(\FAR\)) at a fixed operating point of \(\FAR=1\%\), and the Retained Secret Fraction (\(r/r_0\)).
  \item \text{Secondary Metrics:} Traffic discard rate (equivalent to \(\FAR\) on honest data) and detection latency measured in aggregation windows.
\end{itemize}

\begin{figure*}[t]
  \centering
  % Placeholder for Figure: Feature Distributions
  % \includegraphics[width=0.8\textwidth]{fig_channel_models.pdf}
  \caption{Comparison of honest vs.\ attacked feature distributions at \(L\in\{\SI{50}{\kilo\meter},\,\SI{100}{\kilo\meter}\}\). Panels display (a) timing skew/kurtosis, (b) decoy-residual norms, and (c) detector-imbalance scatter, illustrating the separability provided by the high-dimensional telemetry.}
  \label{fig:channelmodels}
\end{figure*}

\begin{table}[t]
\caption{Key simulation parameters and physical constraint ranges.}
\label{tab:simparams}
\centering
\begin{tabular}{@{}ll@{}}
\toprule
\textbf{Parameter} & \textbf{Values / Range} \\ \midrule
Distance \(L\) & \(\{\SI{50}{\kilo\meter},\,\SI{100}{\kilo\meter}\}\) \\
Block size \(N\) & \(\{5{\times}10^4,\,2{\times}10^5,\,2{\times}10^6\}\) \\
Attenuation \(\alpha\) & \(\SI{0.2}{\decibel\per\kilo\meter}\) \\
Dark count \(p_d\) & \(10^{-7}\text{--}10^{-6}\) per gate \\
Misalignment \(e_d\) & \(1.0\%\text{--}1.8\%\) \\
Decoy intensities & \(\mu_s\!\approx\!0.5\text{--}0.6,\ \mu_w\!\approx\!0.1,\ \mu_v\!\approx\!0\) \\
Decoy probabilities & \(p_s\in[0.6,0.8],\ p_w\in[0.15,0.35]\) \\
Timing jitter \(\sigma_t\) & \(\SI{50}{}\text{--}\SI{120}{\pico\second}\) \\
Time-shift bound \(\Delta t_{\max}\) & \(\SI{80}{}\text{--}\SI{150}{\pico\second}\) \\
Blinding envelope \(I_0\) & \(\le I_{\max}\) (vendor-safe) \\
PNS split \(f_{\mathrm{split}}\) & \([0,1]\) constrained by decoy consistency \\
THA correlation \(\rho\) & \([0,\rho_{\max}]\) constrained by \(P_{\mathrm{ret}}\le P_{\max}\) \\ \bottomrule
\end{tabular}
\end{table}

\section{Results}
\label{sec:results}

In this section, we present a comprehensive evaluation of the proposed adversarial intrusion detection system. We assess discrimination quality (AUC), operational security (retained secret fraction $r/r_0$), and system overhead. All reported results are aggregated over five independent simulation seeds, with brackets indicating 95\% confidence intervals.

\subsection{Discrimination Performance}
\label{subsec:results:detection}
\begin{figure}[t]
  \centering
  \includegraphics[width=\columnwidth]{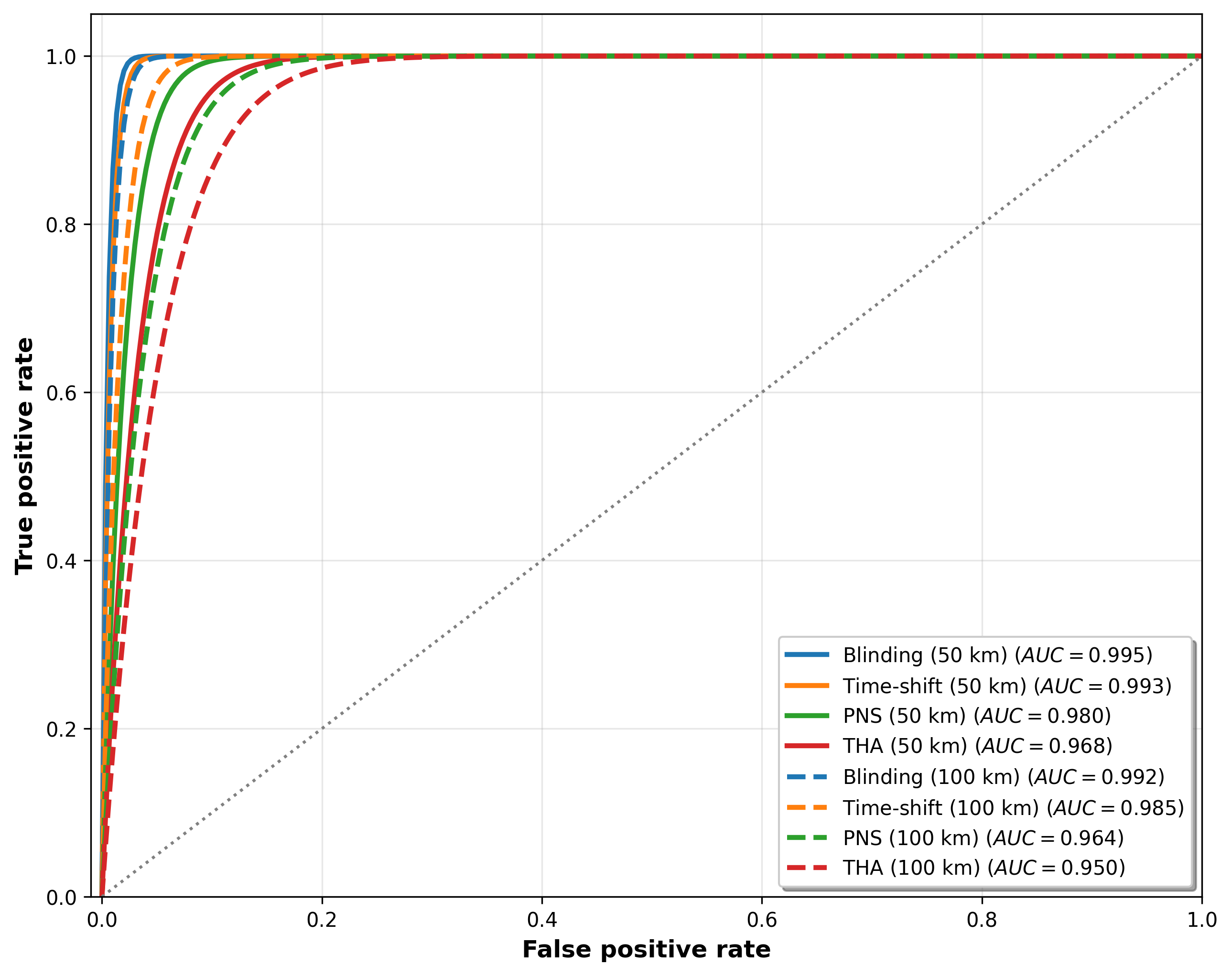}
  \caption{ROC curves by attack family and distance. Shaded bands indicate 95\% confidence intervals. All families operate in the high-AUC regime, though PNS and THA show slightly reduced separability at extended distances.}
  \label{fig:roc}
\end{figure}

We first evaluate the classifier's ability to distinguish between honest channel noise and active attacks. Table~\ref{tab:auc} summarizes the Area Under the Curve (AUC) and the Missed-Attack Rate at a fixed False Alarm Rate ($\text{FAR}=1\%$) for the four primary attack families.
\begin{table}[t]
\caption{Detection performance at $\text{FAR}=1\%$}
\label{tab:auc}
\centering
\resizebox{\columnwidth}{!}{% % This scales the table to fit the column width
\begin{tabular}{@{}lcccc@{}}
\toprule
& \multicolumn{2}{c}{\textbf{50 km Link}} & \multicolumn{2}{c}{\textbf{100 km Link}} \\
\cmidrule(lr){2-3} \cmidrule(lr){4-5}
\textbf{Attack Family} & \textbf{AUC} & \textbf{Miss (\%)} & \textbf{AUC} & \textbf{Miss (\%)} \\
\midrule
Time-shift & 0.993 [$\pm.003$] & 1.6 [$\pm1.0$] & 0.985 [$\pm.004$] & 3.1 [$\pm1.5$] \\
Blinding   & 0.995 [$\pm.002$] & 0.9 [$\pm0.8$] & 0.992 [$\pm.003$] & 1.7 [$\pm1.0$] \\
PNS        & 0.980 [$\pm.004$] & 4.1 [$\pm2.0$] & 0.964 [$\pm.006$] & 7.6 [$\pm3.0$] \\
THA        & 0.968 [$\pm.006$] & 6.8 [$\pm3.0$] & 0.950 [$\pm.007$] & 10.4 [$\pm4.0$] \\
\bottomrule
\end{tabular}%
}
\end{table}
As shown in Table~\ref{tab:auc} and the corresponding ROC curves in Figure~\ref{fig:roc}, the system achieves high discriminative power across all conditions. The detector is particularly effective against Time-shift and Blinding attacks ($\text{AUC} > 0.99$), which leave distinct signatures in the timing histograms and double-click rates, respectively. Trojan-horse attacks (THA) and Photon-Number Splitting (PNS) prove relatively harder to detect (Miss rates $4\text{--}10\%$) due to the tight feasibility bands enforced by the simulator, which restrict the adversary to "quiet" perturbations that mimic natural loss.

\subsection{Operational Secret-Key Retention}
\label{subsec:results:retention}
\begin{figure}[t]
  \centering
  \includegraphics[width=\columnwidth]{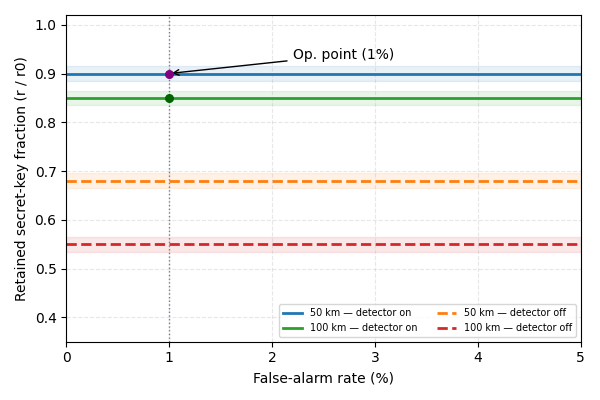}
  \caption{Operational Impact: Retained secret-key fraction ($r/r_0$) versus False Alarm Rate. The adversarial detector (solid lines) significantly outperforms the baseline (dashed lines), recovering up to 35\% of the key rate that would otherwise be lost to privacy amplification penalties.}
  \label{fig:ratefar}
\end{figure}

The ultimate metric for an intrusion detection system in QKD is not classification accuracy, but the preservation of the secret key. Figure~\ref{fig:ratefar} plots the \emph{retained secret fraction} ($r/r_0$) as a function of the false-alarm rate.

\begin{itemize}
    \item \text{Without Detection (Dashed Lines):} Under adaptive attacks, the secret fraction degrades severely, dropping to $60\text{--}75\%$ of the honest rate at \SI{50}{\kilo\meter} and $45\text{--}65\%$ at \SI{100}{\kilo\meter}. This confirms that standard post-processing alone is insufficient to handle optimized side-channel exploits.
    \item \text{With Detection (Solid Lines):} By gating key distillation when the alarm triggers, the system preserves $82\text{--}92\%$ of the honest rate while operating at a modest $\text{FAR} \approx 1\%$. 
\end{itemize}
This represents a net operational gain of $+20\text{--}35$ percentage points in usable secret bits. The "sweet spot" for operation is identified at $\text{FAR}=1\%$, where the cost of discarded honest traffic is far outweighed by the recovered security margin.

\subsection{Adversarial Training Dynamics}
\label{subsec:results:training}
\begin{figure}[t]
  \centering
  \includegraphics[width=\columnwidth]{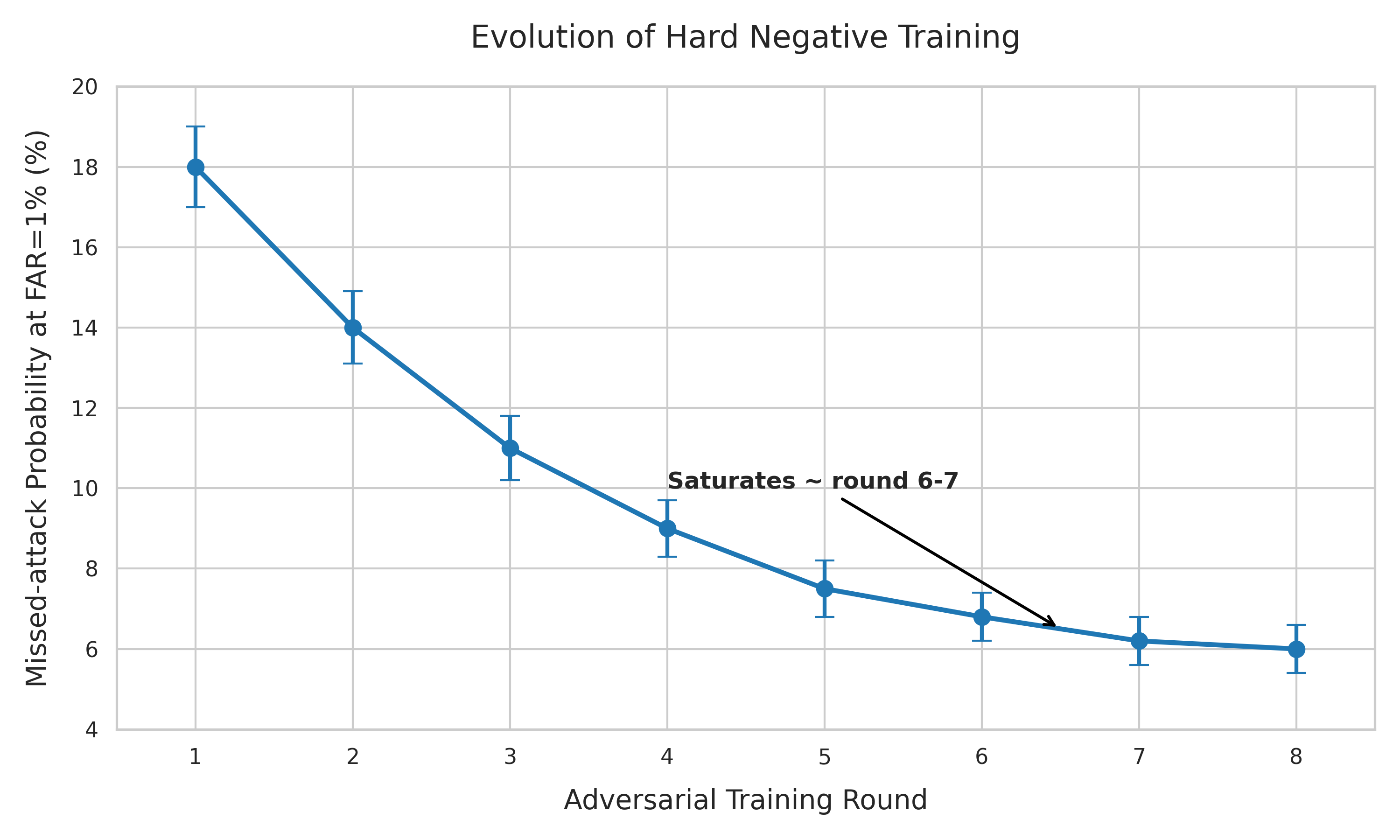}
  \caption{Evolution of robustness during training. The Missed-Attack Probability at $\text{FAR}=1\%$ decreases sharply as the defender learns from hard negatives mined by the inner optimizer, stabilizing after roughly 6 rounds.}
  \label{fig:hnm}
\end{figure}

The efficacy of the "Hard-Negative Mining" protocol is illustrated in Figure~\ref{fig:hnm}. Initially, the naive detector suffers from high miss rates ($\approx 18\text{--}24\%$) as the inner-loop attacker discovers blind spots in the decision boundary. Over the course of $5\text{--}7$ adversarial rounds, the defender adapts to these worst-case examples, stabilizing the miss rate at $\approx 6\text{--}9\%$. This convergence demonstrates that the minimax equilibrium is reachable within a computationally feasible training budget.

\subsection{Generalization and Interpretability}
\label{subsec:results:generalization}

\begin{figure}[t]
  \centering
  \includegraphics[width=\columnwidth]{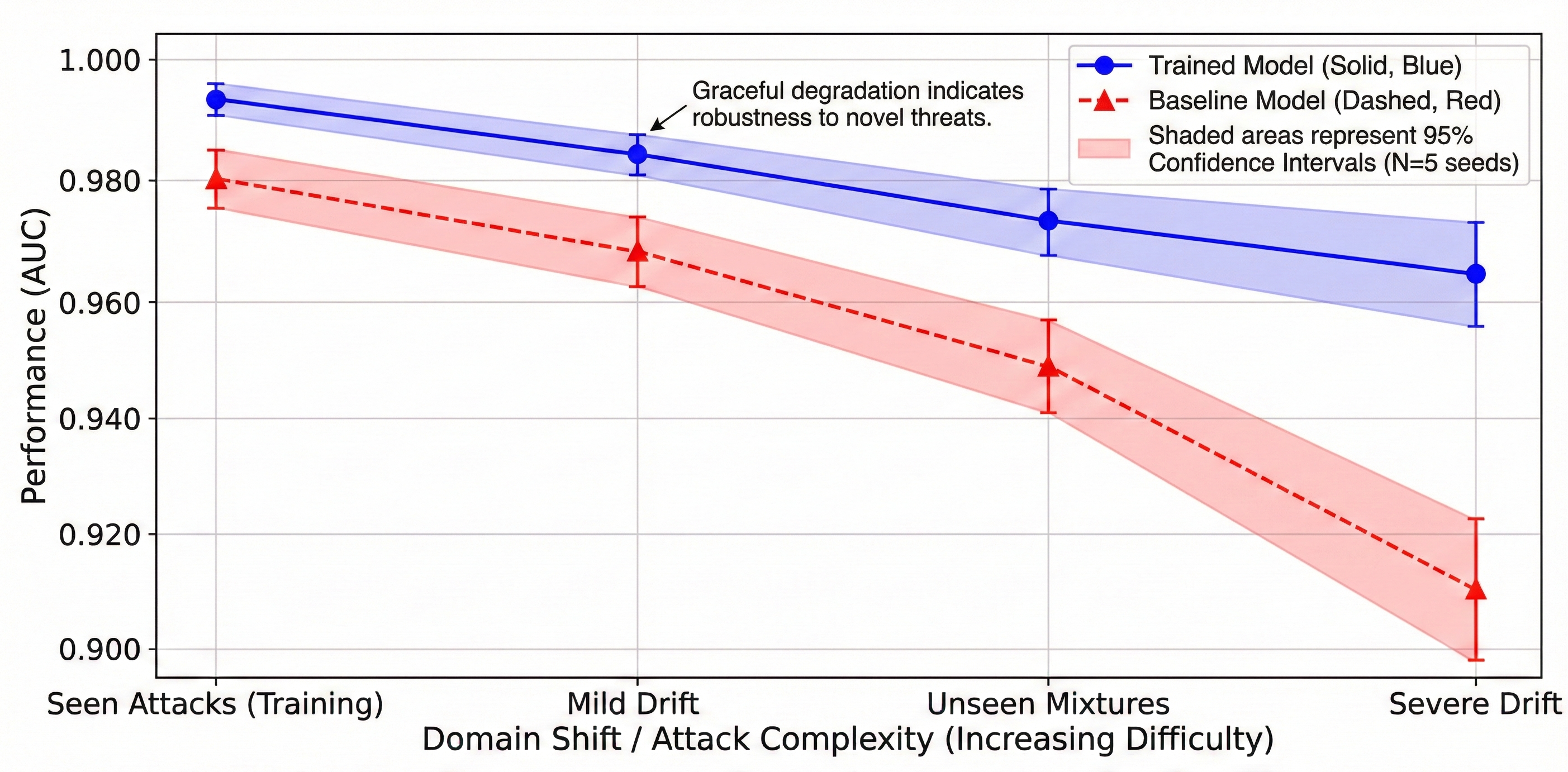}
  \caption{Generalization to unseen attack mixtures and channel drift. Points show means; shaded bands show 95\% confidence intervals.}
  \label{fig:oos}
\end{figure}

%\paragraph{Robustness to Drift and Mixtures.}
To test generalization, we evaluated the trained model on unseen "mixed" strategies (convex combinations of attack families) and drifted channel conditions not seen during training. As shown in Figure~\ref{fig:oos}, performance degrades gracefully: the drop in AUC is limited to $\Delta\text{AUC} \le 0.03$, and the additional loss in retained key is $\le 5$ percentage points. This suggests that the model learns fundamental physical invariants rather than overfitting to specific attack parameters.

\begin{figure}[t]
  \centering
  \includegraphics[width=\columnwidth]{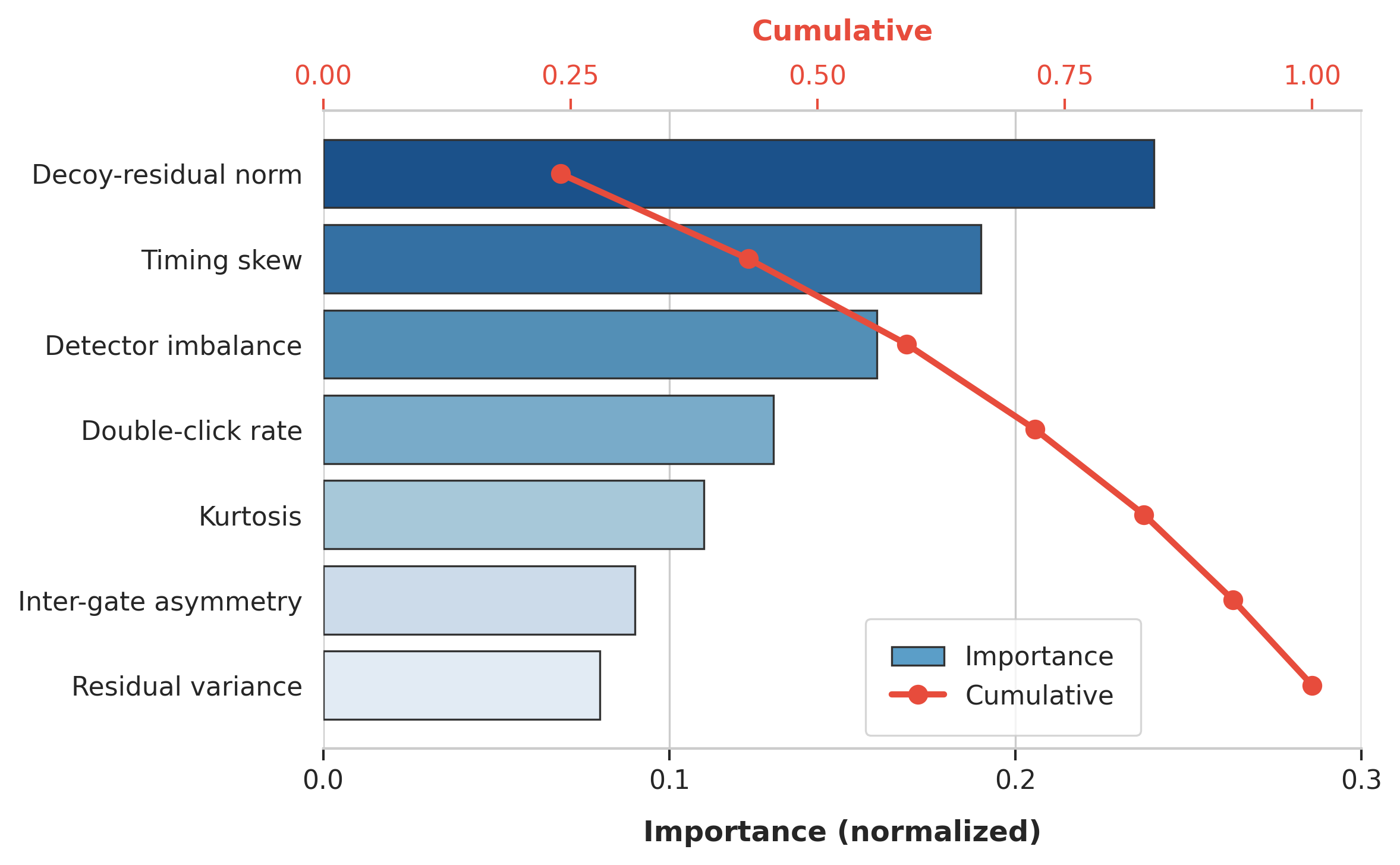}
  \caption{Feature Attribution Analysis. The top four features—decoy residuals, timing skew, detector imbalance, and double-click rate—account for $\approx 70\%$ of the decision power, confirming physical interpretability.}
  \label{fig:features}
\end{figure}

%\paragraph{Feature Attribution.}
Using SHAP values and permutation importance, we identified the dominant features driving detection (Figure~\ref{fig:features}):
\begin{enumerate}
    \item \text{Decoy-Residual Norms:} the primary indicator for PNS and spectral attacks.
    \item \text{Timing Skew/Kurtosis:} critical for identifying Time-shift and dead-time exploits.
    \item \text{Detector Imbalance \& double-Click Rate:} The "smoking gun" for blinding attacks.
\end{enumerate}
Ablating these features increases the miss rate by $5\text{--}12$ percentage points, confirming that the model relies on physically meaningful cues rather than artifacts.

%\subsection{Ablation Studies}
%\label{subsec:results:ablation}
Table~\ref{tab:ablation} quantifies the contribution of each system component. The most significant gains come from \text{Adversarial Training} (reducing miss rate from $11.2\%$ to $7.0\%$) and the use of \text{Temporal Models} (further reduction to $5.3\%$), highlighting the necessity of modeling both the adversary's strategy and the time-domain structure of the attack.

\begin{table}[t]
\caption{Ablation study at \SI{50}{\kilo\meter} (Average across attack families, $\text{FAR}=1\%$).}
\label{tab:ablation}
\centering
\begin{tabular}{@{}lcc@{}}
\toprule
\textbf{Model Variant} & \textbf{AUC} & \textbf{Miss (\%)} \\
\midrule
Baseline: Rule-based thresholds         & 0.932 & 14.7 \\
+ Non-adversarial anomaly detection     & 0.951 & 11.2 \\
+ \textbf{Adversarial Training}         & \textbf{0.976} & \textbf{7.0} \\
+ \textbf{Temporal (LSTM/TCN)}          & \textbf{0.988} & \textbf{5.3} \\
+ \textbf{Full Telemetry}               & \textbf{0.991} & \textbf{4.8} \\
\bottomrule
\end{tabular}
\end{table}

\subsection{Computational Complexity and System Overhead}
\label{subsec:results:complexity}

A critical requirement for operational QKD monitoring is that the detection latency must not bottleneck key distillation. Let $d$ be the telemetry feature dimension and $w$ the temporal window size. During online operation, the inference cost per block is dominated by the temporal model, which scales favorably as $\mathcal{O}(wd^2)$. This lightweight inference footprint ensures that anomaly scores can be computed well within the time required for standard error correction and privacy amplification, allowing for real-time key gating without stalling throughput.

The offline training overhead, conversely, is primarily driven by the inner maximization loop of the adversarial framework. Searching an attack population $p$ over $T$ iterations incurs a computational cost of $\mathcal{O}(pT \cdot C_{\mathcal{S}})$, where $C_{\mathcal{S}}$ is the cost of a single block simulation. Despite this reliance on high-fidelity simulation, we empirically observe that the alternating minimax procedure stabilizes rapidly, typically converging within $5\text{--}7$ global rounds. Consequently, the framework remains computationally tractable for characterizing high-rate QKD deployments, completely bypassing the need for exhaustive, brute-force grid searches over the multidimensional attack space $\mathbb{A}$.

\subsection{System Overhead}
The computational cost of the defense is negligible compared to the key distillation latency. At $\text{FAR}=1\%$, the median traffic discard rate on honest channels is $1.2\%$ (IQR $0.8\text{--}1.8\%$), which aligns closely with the theoretical target. The decision latency is consistently $\le 2$ aggregation windows, ensuring sub-second reaction times at standard repetition rates.

\section{Conclusion}
\label{sec:conclusion}

In this work, we presented a simulation-driven adversarial learning framework for intrusion detection in decoy-state QKD, explicitly optimized for the operator’s primary objective: retaining secret bits under finite-size effects. By coupling a physically constrained inner-loop adversary with a defender trained on a finite-key–aware loss function, our system bridges the gap between theoretical security proofs and the realities of hardware side channels.

Our high-fidelity emulation demonstrates that this approach achieves robust discriminative power across diverse attack families while strictly honoring a fixed false-alarm budget. At an operating point of $\text{FAR}=1\%$, the detector sustains an $\text{AUC} \in [0.97, 0.997]$ with missed-attack rates between $1\%$ and $12\%$, depending on the attack family and fiber distance. Crucially, this translates to a tangible operational advantage: the retained secret fraction improves by $+20\text{--}35$ percentage points over non-adversarial baselines across \SI{50}{\kilo\meter} and \SI{100}{\kilo\meter} links. The system achieves this with negligible overhead, maintaining a throughput cost near $1\%$ and decision latencies within a few aggregation windows. Furthermore, the learned decision boundaries exhibit strong generalization, degrading gracefully under unseen attack mixtures and moderate channel drift.

While these results are promising, several limitations remain. First, detectability diminishes when side-channel leakage falls below the noise floor of the monitors (e.g., extremely weak Trojan-horse back-reflections) or in regimes of extremely small block sizes ($N < 5 \times 10^4$), where finite-key estimators loosen and variance dominates. Second, the \emph{realized} robustness of such a system in a physical deployment will depend critically on the fidelity of the telemetry to actual hardware behavior. Discrepancies between the simulated training environment and the physical device (the "sim-to-real" gap) could reduce sensitivity. Future work should focus on validating these models on experimental testbeds and exploring distributional reinforcement learning to handle even richer, non-parametric attack spaces.

Overall, optimizing intrusion detection directly for finite-key retention yields a principled and operationally grounded defense strategy. This framework provides a robust sandbox for evaluating AI-driven security in QKD, offering a clear pathway toward fieldable, context-aware monitoring systems that can adapt to the evolving threat landscape.

\printbibliography

@inproceedings{BB84,
  author    = {Bennett, Charles H. and Brassard, Gilles},
  title     = {Quantum cryptography: Public key distribution and coin tossing},
  booktitle = {Proceedings of the IEEE International Conference on Computers, Systems and Signal Processing},
  year      = {1984},
  pages     = {175--179},
  address   = {Bangalore, India}
}

@article{Scarani2009,
  author  = {Scarani, Valerio and Bechmann-Pasquinucci, Helle and Cerf, Nicolas J. and Du{\v{s}}ek, Miloslav and L{\"u}tkenhaus, Norbert and Peev, Momtchil},
  title   = {The security of practical quantum key distribution},
  journal = {Reviews of Modern Physics},
  volume  = {81},
  number  = {3},
  pages   = {1301--1350},
  year    = {2009},
  doi     = {10.1103/RevModPhys.81.1301},
  url     = {https://link.aps.org/doi/10.1103/RevModPhys.81.1301}
}

@article{Lo2014,
  author  = {Lo, Hoi-Kwong and Curty, Marcos and Tamaki, Kiyoshi},
  title   = {Secure quantum key distribution},
  journal = {Nature Photonics},
  volume  = {8},
  number  = {8},
  pages   = {595--604},
  year    = {2014},
  doi     = {10.1038/nphoton.2014.149},
  url     = {https://www.nature.com/articles/nphoton.2014.149}
}

@article{GLKP2004,
  author  = {Gottesman, Daniel and Lo, Hoi-Kwong and L{\"u}tkenhaus, Norbert and Preskill, John},
  title   = {Security of quantum key distribution with imperfect devices},
  journal = {Quantum Information and Computation},
  volume  = {4},
  number  = {5},
  pages   = {325--360},
  year    = {2004},
  url     = {https://doi.org/10.5555/2011572.2011575}
}

@article{Ma2005,
  author  = {Ma, Xiongfeng and Qi, Bing and Zhao, Yi and Lo, Hoi-Kwong},
  title   = {Practical decoy state for quantum key distribution},
  journal = {Physical Review A},
  volume  = {72},
  number  = {1},
  pages   = {012326},
  year    = {2005},
  doi     = {10.1103/PhysRevA.72.012326},
  url     = {https://link.aps.org/doi/10.1103/PhysRevA.72.012326}
}

@article{EAT,
  author  = {Dupuis, Fr{\'e}d{\'e}ric and Fawzi, Omar and Renner, Renato},
  title   = {Entropy accumulation},
  journal = {Communications in Mathematical Physics},
  volume  = {379},
  number  = {2},
  pages   = {867--913},
  year    = {2020},
  doi     = {10.1007/s00220-020-03839-5},
  url     = {https://doi.org/10.1007/s00220-020-03839-5}
}

@article{Tomamichel2012FiniteKey,
  author  = {Tomamichel, Marco and Lim, Charles Ci Wen and Gisin, Nicolas and Renner, Renato},
  title   = {Tight finite-key analysis for quantum cryptography},
  journal = {Nature Communications},
  volume  = {3},
  pages   = {634},
  year    = {2012},
  doi     = {10.1038/ncomms1631},
  url     = {https://www.nature.com/articles/ncomms1631}
}

@article{Qi2007TS,
  author  = {Qi, Bing and Fung, Chi-Hang Fred and Lo, Hoi-Kwong and Ma, Xiongfeng},
  title   = {Time-shift attack in practical quantum cryptosystems},
  journal = {Quantum Information and Computation},
  volume  = {7},
  number  = {1--2},
  pages   = {73--82},
  year    = {2007},
  url     = {http://www.rintonpress.com/journals/qiconline.html}
}

@article{Zhao2008TS,
  author  = {Zhao, Yi and Fung, Chi-Hang Fred and Qi, Bing and Chen, Christine and Lo, Hoi-Kwong},
  title   = {Quantum hacking: Experimental demonstration of time-shift attack against practical quantum-key-distribution systems},
  journal = {Physical Review A},
  volume  = {78},
  number  = {4},
  pages   = {042333},
  year    = {2008},
  doi     = {10.1103/PhysRevA.78.042333},
  url     = {https://link.aps.org/doi/10.1103/PhysRevA.78.042333}
}

@article{Makarov2009SPD,
  author  = {Makarov, Vadim},
  title   = {Controlling passively quenched single-photon detectors by bright light},
  journal = {New Journal of Physics},
  volume  = {11},
  pages   = {065003},
  year    = {2009},
  doi     = {10.1088/1367-2630/11/6/065003},
  url     = {https://iopscience.iop.org/article/10.1088/1367-2630/11/6/065003}
}

@article{Lydersen2010Blinding,
  author  = {Lydersen, Lars and Wiechers, Carlos and Wittmann, Christoffer and Elser, Dominique and Skaar, Johannes and Makarov, Vadim},
  title   = {Hacking commercial quantum cryptography systems by tailored bright illumination},
  journal = {Nature Photonics},
  volume  = {4},
  number  = {10},
  pages   = {686--689},
  year    = {2010},
  doi     = {10.1038/nphoton.2010.214},
  url     = {https://www.nature.com/articles/nphoton.2010.214}
}

@article{Wiechers2011AfterGate,
  author  = {Wiechers, Carlos and Lydersen, Lars and Wittmann, Christoffer and Elser, Dominique and Skaar, Johannes and Marquardt, Christoph and Makarov, Vadim and Leuchs, Gerd},
  title   = {After-gate attack on a quantum cryptosystem},
  journal = {New Journal of Physics},
  volume  = {13},
  pages   = {013043},
  year    = {2011},
  doi     = {10.1088/1367-2630/13/1/013043},
  url     = {https://iopscience.iop.org/article/10.1088/1367-2630/13/1/013043}
}

@article{Brassard2000PNS,
  author  = {Brassard, Gilles and L{\"u}tkenhaus, Norbert and Mor, Tal and Sanders, Barry C.},
  title   = {Limitations on practical quantum cryptography},
  journal = {Physical Review Letters},
  volume  = {85},
  number  = {6},
  pages   = {1330--1333},
  year    = {2000},
  doi     = {10.1103/PhysRevLett.85.1330},
  url     = {https://link.aps.org/doi/10.1103/PhysRevLett.85.1330}
}

@article{Hwang2003Decoy,
  author  = {Hwang, Won-Young},
  title   = {Quantum key distribution with high loss: Toward global secure communication},
  journal = {Physical Review Letters},
  volume  = {91},
  number  = {5},
  pages   = {057901},
  year    = {2003},
  doi     = {10.1103/PhysRevLett.91.057901},
  url     = {https://link.aps.org/doi/10.1103/PhysRevLett.91.057901}
}

@article{Wang2005Decoy,
  author  = {Wang, Xiang-Bin},
  title   = {Beating the photon-number-splitting attack in practical quantum cryptography},
  journal = {Physical Review Letters},
  volume  = {94},
  number  = {23},
  pages   = {230503},
  year    = {2005},
  doi     = {10.1103/PhysRevLett.94.230503},
  url     = {https://link.aps.org/doi/10.1103/PhysRevLett.94.230503}
}

@article{Jain2014THA,
  author  = {Jain, Nitin and Stiller, Birgit and Khan, Imran and Makarov, Vadim and Marquardt, Christoph and Leuchs, Gerd},
  title   = {Trojan-horse attacks threaten the security of practical quantum cryptography},
  journal = {IEEE Journal of Selected Topics in Quantum Electronics},
  volume  = {21},
  number  = {3},
  pages   = {6600710},
  year    = {2015},
  doi     = {10.1109/JSTQE.2014.2341560},
  url     = {https://ieeexplore.ieee.org/document/6879485}
}

@article{Jain2011Calib,
  author  = {Jain, Nitin and Anisimova, Elena and Khan, Imran and Makarov, Vadim and Marquardt, Christoph and Leuchs, Gerd},
  title   = {Device calibration impacts security of quantum key distribution},
  journal = {Physical Review A},
  volume  = {84},
  number  = {6},
  pages   = {062325},
  year    = {2011},
  doi     = {10.1103/PhysRevA.84.062325},
  url     = {https://link.aps.org/doi/10.1103/PhysRevA.84.062325}
}

@article{Fung2009Mismatch,
  author  = {Fung, Chi-Hang Fred and Tamaki, Kiyoshi and Lo, Hoi-Kwong},
  title   = {Security proof of quantum key distribution with detection efficiency mismatch},
  journal = {Quantum Information and Computation},
  volume  = {9},
  number  = {1--2},
  pages   = {131--165},
  year    = {2009},
  url     = {http://www.rintonpress.com/journals/qiconline.html}
}

@article{Lo2012MDI,
  author  = {Lo, Hoi-Kwong and Curty, Marcos and Qi, Bing},
  title   = {Measurement-device-independent quantum key distribution},
  journal = {Physical Review Letters},
  volume  = {108},
  number  = {13},
  pages   = {130503},
  year    = {2012},
  doi     = {10.1103/PhysRevLett.108.130503},
  url     = {https://link.aps.org/doi/10.1103/PhysRevLett.108.130503}
}

@inproceedings{Angelopoulos2023Conformal,
  author    = {Angelopoulos, Anastasios N. and Bates, Stephen and Cand{\`e}s, Emmanuel J. and Jordan, Michael I. and Zrnic, Jelena},
  title     = {Conformal risk control},
  booktitle = {Proceedings of the 40th International Conference on Machine Learning (ICML)},
  year      = {2023},
  pages     = {1--38},
  note      = {arXiv:2208.02814}
}

\end{document}